\documentclass[10pt,a4paper,twoside]{article}
\usepackage{epsfig}
\usepackage{baltlat6}
\usepackage{array}
\usepackage{here}
\begin{document}
\ \
\vspace{0.5mm}
\setcounter{page}{315}

\titlehead{Baltic Astronomy, vol.\,22, 315--328, 2013}

\titleb{PHYSICAL\,\,ALTERNATIVE\,\,TO\,\,THE\,DARK\,\,ENERGY\,\,PARADIGM}

\begin{authorl}
\authorb{A. Sapar}{}
\end{authorl}
\begin{addressl}
\addressb{}{Tartu Observatory, 61602 T\~{o}ravere, Estonia; sapar@to.ee}
\end{addressl}
\submitb{Received: 2013 October 9; revised October 22; accepted: November 6}

\begin{summary} The physical nature of the presently dominating enigmatic dark
energy in the expanding universe is demonstrated to be explainable as an excess
of the kinetic energy with respect to its potential energy. According to
traditional Friedman cosmology, any non-zero value of the total energy integral
is ascribed to the space curvature. However, as we show, in the flat universe
the total energy also can be different from zero. Initially, a very small excess
of kinetic energy originates from the early universe. The present observational
data show that our universe has probably a flat space with an excess of kinetic
energy. The evolutionary scenario shows that the universe presently is in the
transitional stage where its radial coordinate expansion approaches the velocity
of light. A possibility of the closed Bubble universe with the local Big Bang
and everlasting expansion is demonstrated. Dark matter can be essentially
contributed by the non-relativistic massive neutrinos, which have cooled to very
low temperatures and velocities thus favoring the formation of the observed
broad equipotential wells in galaxies.
\end{summary}


%% \resthead is the RUNNING TITLE at top of the pages
\resthead{Physical alternative to the dark energy paradigm}
{A. Sapar}

\sectionb{1}{INTRODUCTION}

During the last decades a strange situation in fundamental cosmology
has arisen. On the observational side, a considerable progress has
been made, with the results appreciated by awarding in 2011 the
Nobel Prize to S. Perlmutter, A. Riess and B. P. Schmidt. In their
Nobel Lectures, these eminent scientists ad- dressed theorists with
an appeal to unveil the physical nature of the mysterious dark
energy. A large number of papers have been devoted to the problem,
but, unfortunately, without a considerable success.

This circumstance impelled us to propose a physically clear and simple alter-
native concept of the nature of dark energy. Such a possibility was discussed by
us about a half-century ago (Sapar 1964). In order to explain the problem in more
details, it is necessary to turn directly to the Einstein equations of general
relativity. As it was emphasized by Albert Einstein himself, the concept of the curved
space-time expressed in the curvature tensor is elegant and has a deep physical
nature. Einstein perceived that the energy-momentum tensor, equalized to his
elegant tensor of space-time geometry, is not perfect. The problems, connected with
the meaning and necessity of the cosmological constant in the energy-momentum
tensor, are also known. From a new standpoint we try to interpret the recent
observational data connected with (1) the Ia supernovae (hereafter SNe) as standard
candles for cosmology, and (2) the results of the Antarctic high-balloon Boomerang
mission which has investigated the fine structure of the cosmic microwave
background (CMB) angular power spectrum and has added important contribution
to the constraints on the variety of different theoretically assumed forms of the
energy-momentum tensor. The results of the Boomerang mission have added an
important constraint, that the space in cosmological dimensions with high fidelity
turns out to be flat.

The study of SNe has shown that the expansion of the universe exceeds
somewhat the expected expansion rate, corresponding to the accepted cosmological
standard model with the observed matter, dark matter and dark energy. The
result has been predominantly treated as mysterious, which demands modification
of the Friedman equations. Different additional hypothetical source-terms were
added to these equations, and presently there is a wide variety of proposed
model-equations for the universe. The best known of them add the quintessence term
or the repulsion term of the cosmological constant, which gives an exponential
expansion of the universe in the future.

During the last decade, the intensive observational studies of the law and
nature of the expansion of the universe have been accomplished with the large
ground-based and orbital telescopes, treating the light curves of SNe as standard
candles. The main result of these investigations -- the accelerating expansion of
the flat-space universe containing about 70\% of enigmatic dark energy -- has been
generally acknowledged as a paradigm by the main-stream cosmologists. An early
paper with the results of the SNe project was published by Goldhaber \&
Perlmutter (1998). Important results were also obtained by the international mission
Supernova Legacy Survey (SLS), based on the dependence 'redshift versus ob-
served stellar magnitude' of SNe. A detailed analysis of the methods applied in
the study and a review of the SLS mission results were published by Astier et al.
(2006). The paper has 42 authors from 18 institutions of 7 countries. Leibundgut
(2009) emphasized that with the SLS project a new era in observational cosmology
has been started. In above mentioned papers and the Nobel lectures the project
leaders appeal to theorists for their help in the unraveling the physical nature of
dark energy. Summarizing the results of the Boomerang mission, considering the
angular structure of CMB, MacTavish et al. (2006) show, that our universe is very
close to being flat.

\sectionb{2}{PHYSICAL AND HISTORICAL BACKGROUND}

The situation described above seems to be too complicated, and we prefer
to start from the conservative Occam's razor principle, which suggests to avoid
introducing any new parameters of unknown nature without the extreme need.
The present situation in fundamental cosmology is strange in the aspect that the
total energy integral in the flat-space universe has been assumed to be strictly
zero. This seems to be a physically groundless constraint. From such a treatment
of the usual Friedman equations follows, that even a minimal non-zero value of
the total energy in the universe is inevitably related to the space curvature. Such
a point of view seems to be unacceptable.

In order to make a step forward in our introductory discussion, we mention
that the main additional assumptions to be added to the Friedman equations are
the matter state equations. This is a crucial point in the interpretation and study
of the Friedman equations. As we have shown many years ago (Sapar 1964) in
Publications of Tartu Observatory (written in Russian, but with a quite detailed
English summary), an important class of special Friedman equations of state can
be given by
\begin{equation}
G\rho_nR^n=C_n,~~~~~~
(n-3)\rho_n=3 P_nc^{-2}.
\end{equation}
Here $G$ is the gravitational constant, $\rho_n$ and  $P_n$ are the partial densities
of matter and pressure, and $C_n $ are the constants.
Every such component can be considered as a contribution to the equation of
total energy and total pressure in the equations of the evolution of the
universe. As we see, the case $n=3$ corresponds to the usual non-relativistic matter,
being at rest in the co-moving cosmological coordinate framework, and the case $n=4$
corresponds to the radiation, which is redshifted due to expansion of the
universe. Most of the present cosmologists write the equation
in the form
\begin{equation}
p_n\propto w_n\rho_n,
\end{equation}
where the index $w_n$, which we name the pressure index, is given by
\begin{equation}
w_n={n-3\over 3}.
\label{ww}
\end{equation}
Thus, $w_n=0$ for the classical matter and $w_n=1/3$ for the radiation. In the evolving
universe the index changes from 1/3 to 0. The equations, which specify such an evolutionary
change, have been published recently (Sapar 2011).

If in the evolving universe there are several contributors to the matter density
which interact only gravitationally, then for the total density we can write:
\begin{equation}
G\rho=\sum_n {C_n\over R^n }.
\label{rhon}
\end{equation}
For the pressure we obtain in a similar way the equation
\begin{equation}
GP=c^2\sum_n w_n {C_n\over R^n }.
\label{pren}
\end{equation}
Now let us check which values of the parameter $n$ are urgently
needed in the present context. First, it is inevitable to remove the
constraint that in the flat space only the zero value of total
energy is possible. Thus, we start from the Einstein equation in the
form, given in our earlier paper (Sapar 1964, Eq. 3.114), but
repeated here in somewhat differently scaled form:
\begin{equation}
{\dot R^2\over c^2}= {\alpha \over R}+{\beta \over R^2}+\kappa-k.
\label{R2}
\end{equation}
The non-zero positive constant $\kappa$ is the constant of energy, corresponding to the
energy-momentum tensor, which can also be non-zero in the flat space, where the
curvature constant $k=0$. The constant $\kappa$ can be also added to the cases $k=\pm1$.
This means that, in addition to the rest-matter and radiation, we need also the
term corresponding to $n=2$. This term is a very natural integral of energy in
the flat space too. This energy might be very small and unnoticeable parameter
during the early evolutionary stages of the universe, like the asymmetry of the
matter and antimatter contributions.

The model universes with $\kappa>0$ we name for shortness the  kinetic
energy dominated (KED) model universes.

As is evident from the given equations, the contribution by the energy integral
in any universe, including the flat model universe, gives the contribution in the
form of negative pressure, in this case:
\begin{equation}
P_2=-\rho_2{c^2\over 3}.
\label{n2}
\end{equation}

Thus, the non-localized integral value of energy can  formally be localized, using the term with
negative pressure. Thus, the enigmatic accelerating contribution to the expansion of the universe
can be attributed simply not to dark energy, but to the excess of the
kinetic energy of matter  in our  expanding KED universe, which at the present epoch  gives about 70\%
of the contribution to the Hubble expansion rate. During early evolutionary stages of the universe,
i.e. at the small values of $R$, this contribution was unnoticeable. The term 'negative pressure' is due to
the incorrect interpretation of the term of non-zero total energy integral, reduced
to the terms of local equation of state. The used partial contributions can be
compared to the Fourier series expansions, where the components do not always
have direct physical interpretation.

In order to get a better {\it Vorstellung} about the role of constant term $\kappa -k$
in Equation (\ref{R2}),
we study different possibilities of the interpretation
of cosmological scenario in the present era,
considering separately the scenarios of the evolution of the universe
if it has  the flat space, the hyperbolic space
or the elliptical space.
All of these  can be treated by the  KED model universes.

In the present state of investigations there is a possibility, which cannot be
completely excluded, of abandoning the presumption of the flatness of the universe
and accepting the possibility that the universe is hyperbolic. However, both model
universes, flat and hyperbolic, have the shortcoming that the universe originates
from a self-creational act in the infinite space. In this case, any localization of
the Big Bang, even at the very beginning, is excluded. Thus, both the flat and
the hyperbolic $k=-1$ universe models have this principal defect in common.
Fortunately, there is also a third and principally elegant possibility of the elliptical
Bubble universe, which we discuss later in this paper.

\sectionb{3}{HYPERBOLIC UNIVERSE WITH INTEGRAL ENERGY AS DARK\\ ENERGY}

Let us now start to study different model universes. Many years ago we derived
analytical solutions for uniform model universes filled with matter and radiation,
for the event (past) and future (particle) light cones therein, for different distances,
observables and distributions (Sapar 1964, 1965, 1966, 1970, 1976). Recently
(Sapar 2011) we have started to actualize some of the results in the context of
progress in observational cosmological studies during the last decades, in particular
in connection with the exciting results of studies of the SNe as standard light
sources at cosmological redshifts.

In the present section we investigate, which is the evolutionary scenario of the
universe if we accept the concept of a hyperbolic universe instead of dark energy in
the flat-space universe. For the uniform (isotropic and homogeneous) hyperbolic
universe we obtain from Equation (\ref{R2}) in the isotropic co-moving reference frame the equation
\begin{equation}
{\dot R^2\over c^2}= {\alpha\over R}+{\beta \over R^2}+1.
\label{dR2dt}
\end{equation}
Here $\alpha$ corresponds to the gravitational potential of the matter in
the co-moving coordinate frame,
and $\beta $ is the gravitational potential term, corresponding to relativistic particles and radiation.
The last term in this equation, treated in the spirit of the Mach principle, is the gravitational potential
of the universe.
 This potential energy of the universe equals to $c^2 $.
 The constants used here and in Equation (\ref{R2}) are defined by
 \begin{equation}
\alpha={2GM\over c^2},~~~~~~M={4\pi\rho_m R^3\over 3},
\end{equation}
\begin{equation}
\beta={2GP\over c^2},~~~~~~P={4\pi\rho_r R^4\over 3}.
\end{equation}

In these equations $\rho_m$ and  $\rho_r$ are the density of matter and the
density of radiation, respectively.
From  Equation  (\ref{dR2dt}) we see that finally the expansion rate of the universe, $\dot R$,
differently from the scenario with the cosmological constant,  tends to $c$.
As was shown in Sapar (1964, 2011), the integrating gives for the age of the hyperbolic
 universe an analytical expression
\begin{equation}
 ct=Q-{\alpha\over 2}\omega, ~~~~~Q={R\dot R\over  c}=\sqrt{ R^2+\alpha R+\beta}.
 \label{ct}
\end{equation}
Here $\omega$ corresponds to the equation of the  light cone, which carries us astronomical information
 of the past events, which locally is defined as the zero-geodetic by
\begin{equation}
c^2dt^2=R^2d\omega^2={R_0^2\over (1+z)^2}d\omega^2.
\end{equation}
The angular distance $\omega$ on the light cone is given  by
\begin{equation}
\omega= c\int{dt\over R}= c\int{dR\over R\dot R}={\rm ln}y,
\end{equation}
where (Sapar 1964)
\begin{equation}
y= 2Q+{2R+\alpha}.
\label{dRdtL}
\end{equation}
The light flux at the geometrical
distance  $D_g$ is
\begin{equation}
 F_g={L\over S}={L\over 4\pi D_g^2},~~~~D_g=  R_0\omega_z.
\end{equation}
The luminosity distance $D_L$ is given by the
 observational flux, $F_L$, defined by
\begin{equation}
F_L={L\over S_L}={L\over 4\pi D_L^2},
\end{equation}
where the luminosity distance
\begin{equation}
D_L=( 1+z)D_g=( 1+z)R_0\sinh\omega_z,
\end{equation}
and the angular distance
\begin{equation}
\omega_z=\ln\bigg({y_0\over y_z}\bigg).
\end{equation}
Thus, we obtain
\begin{equation}
\sinh\omega_z={1\over 2}\bigg({y_0\over y_z}- {y_z\over y_0}\bigg).
\end{equation}

These equations specify the observed brightness values of distant cosmological
standard candles, SNe, in a hyperbolic universe. Instead of dark energy the model
incorporates the usual energy integral in the hyperbolic space, corresponding to the
excess of kinetic energy compared to the potential energy. However, it is necessary
to demonstrate that the observed expansion of the universe, at least the results
of studies of SNe, can be treated in terms of an expanding hyperbolic universe.
Adequacy of such a universe is strongly restricted by the results of cosmological
triangulation of the parallax by the Boomerang missions, from which it has been
found that with quite high probability the cosmological space of the universe is
flat.

The present value of the radial coordinate, $R_0$ , for a hyperbolic universe is
specified from Equation (\ref{dR2dt}) uniquely by
\begin{equation}
R_0={c\over H_0}\bigg({\alpha\over R_0}+{\beta\over R_0^2}+1\bigg)^{1/2},
\end{equation}
where $H_0$ is the present Hubble constant, $H_0=\dot R_0/R_0$.

\sectionb{4}{THE FLAT UNIVERSE WITH AN EXCESS OF KINETIC ENERGY}

Here we will consider the flat model universe with an excess of kinetic energy,
ignoring the presence of any dark energy (a KED universe). In this case Equation
(\ref{R2}) reduces to the form
\begin{equation}
{\dot R^2\over c^2}= {\alpha \over R}+{\beta \over R^2}+{\kappa}.
\end{equation}
Now we calibrate here the values of $R$ so that $\kappa=1$. In this way we obtain  the
 equations of the same form as for a hyperbolic universe.
The present value of the Hubble constant gives a constraint
\begin{equation}
{H_0^2}= {\alpha c^2\over R_0^3}+{\beta c^2\over R_0^4}+{ c^2\over R_0^2}
\end{equation}
between $R_0$,  $\alpha$ and $\beta$   for the flat KED  universe.
To simplify the equations, we take into use the redshift  factors $\tau=1+z$, obtaining
\begin{equation}
R={R_0\over \tau}.
\end{equation}
For the Hubble constant at any redshift now we can write
\begin{equation}
{H^2}= {\alpha \tau^3 c^2\over R_0^3}+{\beta \tau^4c^2\over R_0^4}+{\tau^2 \over R_0^2}.
\end{equation}

For the angular distance from the light cone we obtain the same general
expression as for a hyperbolic universe, namely,
\begin{equation}
\omega_z= c\int\limits_t^{t_0}{dt\over R}= c\int\limits_R^{R_0}{dR\over R\dot R}=
c\int\limits_1^{1+z}{ d\tau \over \tau \dot R}.
\end{equation}

Using the Hubble constant, in a similar way as in the case of the hyperbolic
space, we obtain
\begin{equation}
\omega_z= {c\over R_0} \int\limits_1^{\tau}{ d\tau\over  H}=\ln\bigg({y_0\over y_z}\bigg),
\end{equation}
where  $y_z$ is given by Equation (\ref{dRdtL}).

The luminosity distance in the case of the flat universe is different from that
in the hyperbolic space. Namely, in the flat space
\begin{equation}
D_L=R_0(1+z)\omega_z=R_0(1+z)\ln\bigg({y_0\over y_z}\bigg).
\end{equation}
The equations for the Hubble constant and for the luminosity distance are use-
ful for computing the 'stellar magnitude versus redshift' dependence and for its
comparison with the observational brightness values of SNe.

The ratio of the luminosity distances for the hyperbolic and for the flat universe
with the kinetic energy excess is given by
\begin{equation}
r_z={\sinh(\omega_z)\over \omega_z },~~~~~~~~\omega_z=\ln\bigg({y_0\over y_z}\bigg).
\label{rz}
\end{equation}
This expression means that the luminosity distance in the hyperbolic space is always larger than in the
flat space.
From here for the difference of the observed stellar magnitudes we obtain
\begin{equation}
\Delta m_z=5 \log r_z.
\end{equation}
Now we estimate the ratio in the case of presently accepted value of $\alpha$.
For small values of $\omega_z$ we can write
\begin{equation}
r_z=1+{\omega_z^2\over 6}.
\end{equation}
At $z=1$ we obtain $y_0/y_z\approx1.67$ and thus, $w_z\approx0.5$, from where $r_z\approx1.044$.
In stellar magnitudes this means that
\begin{equation}
\Delta m_z=5 \log 1.044\approx0.1.
\end{equation}
This difference is small, and  there is no
need to make presently the detailed computations both for the curved space
and for the flat KED universe.

For the age of the universe we have now
\begin{equation}
ct={R_0\dot R_0\over c}-{\alpha\Omega_0\over 2},
\end{equation}
 where the angular distance of the past events horizon
\begin{equation}
\Omega_0=\ln\bigg({2R_0\dot R_0\over c \alpha}\bigg).
\end{equation}
The differences between different model universes diminish in the direction of redshifts.
As it has been shown by Aldering et al. (2007), at $1.7<z<3$ the evolutionary differences
between  the model universes give additional differences in apparent stellar magnitudes, which are less than $\Delta m=0.02$.

Here deserves mentioning that in theoretical cosmology there are two serious
studies generalizing the redshift versus magnitude equations,
Kaufman \& Schucking (1971) and Kaufman (1971), both very detailed and on high mathematical
level. The results of these papers were generalized using the Weierstrass elliptical
functions by Darbowsky \& Stelmach (1986).

\sectionb{5}{THE FLAT UNIVERSE WITH COSMOLOGICAL CONSTANT AS\\ DARK ENERGY}

A similar analysis of the equations can be carried out for the
flat model universe with the cosmological constant $\Lambda$, as a  version to treat the dark energy.
In this case we  start from the equation
\begin{equation}
{\dot R^2}= {\alpha c^2\over R}+{\beta c^2\over R^2}+{\Lambda R^2\over 3}.
\end{equation}
From here we see, that $\Lambda$ corresponds to $n=0$, or $w_n=-1$.
Thus, we can find the  value of $\Lambda$ corresponding to observations from
\begin{equation}
{H_0^2}= {\alpha c^2\over R_0^3}+{\beta c^2\over R_0^4}+{\Lambda \over 3}.
\end{equation}
This equation is a constraint between  the values of $R_0$,
 $\alpha$, $\beta$ and $\Lambda$.

To obtain simpler equations for computations, similarly to  the case of the flat KED universe,
 we take  into use the redshift factors $\tau=1+z$.
For the light cone and for the age of such model universe it is necessary
to carry out the numerical computations, because the analytical equation is lacking.
This holds also for finding of the dependence 'stellar magnitude versus redshift' to
check whether it corresponds to the  observational brightness-curve
 of SNe at  large cosmological distances.\\

\sectionb{6}{COMPARISON OF THE HYPERBOLIC AND FLAT MODEL UNIVERSES}

We have carried out the computations both for the angular variable $\omega$  and the age of the
universe using the value $H_0=73$ km s$^{-1}$ Mpc$^{-1}$ for the present Hubble constant.
The corresponding present  critical mass density  is
\begin{equation}
\rho_{cr}= {3H_0^2\over 8\pi G}=1.0010\cdot10^{-29}~~~~~~[{\rm g/cm^3}]
\end{equation}
and the corresponding Hubble age of the universe is $H_0^{-1}=13.39 $  Gyr.

\begin{figure}[!tH]
\vbox{
\centerline{\psfig{figure=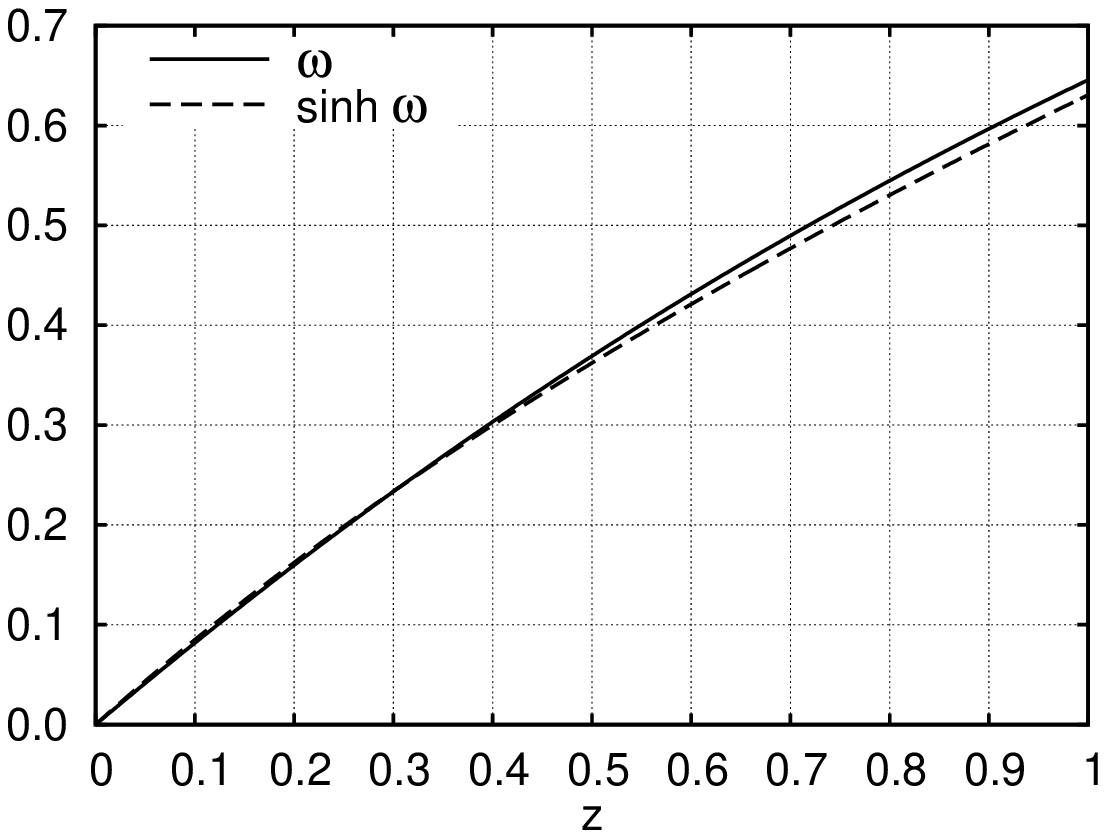,width=100mm,angle=0,clip=}}
\vspace{1mm}
\captionb{1a}
{The run of the angular variable $\omega$ for the flat model universe with cosmological
 constant (solid curve) and $\sinh\omega$ for a hyperbolic universe (dashed curve) versus the redshift. The difference of curves is rather small.}
}
\end{figure}
\begin{figure}[!tH]
\vbox{\vspace{-5mm}
\centerline{\psfig{figure=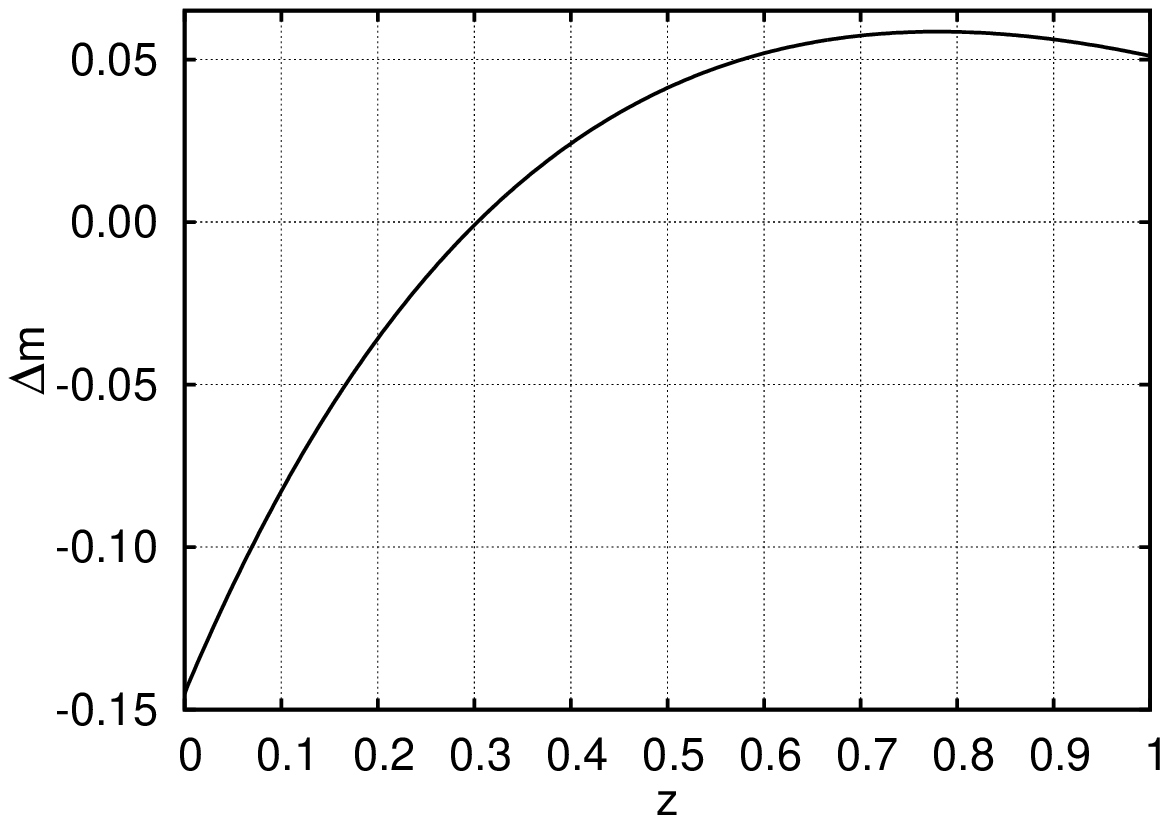,width=100mm,angle=0,clip=}}
\vspace{1mm}
\captionb{1b}
{The ratio of the luminosity distances for the above described flat and hyperbolic model universes.
The difference is quite small and apparently not distinguishable by  observations.}
}
\end{figure}

We checked different densities of matter to study acceptability of the hyperbolic
space for explication of the results
of the apparent luminosity versus redshift curves. These curves are treated usually
 as testimony of urgent need for the  cosmological
constant in the flat universe. The results suggested to us that it is possible to explain
the dependence alternatively, treating the dark energy as an   integral of total energy.
It corresponds to a very small initial excess of the kinetic energy relative to the potential energy
of the matter and radiation in the primordial universe.
This bias is similar to a small primordial matter and antimatter
asymmetry.

\begin{figure}[!tH]
\vbox{
\centerline{\psfig{figure=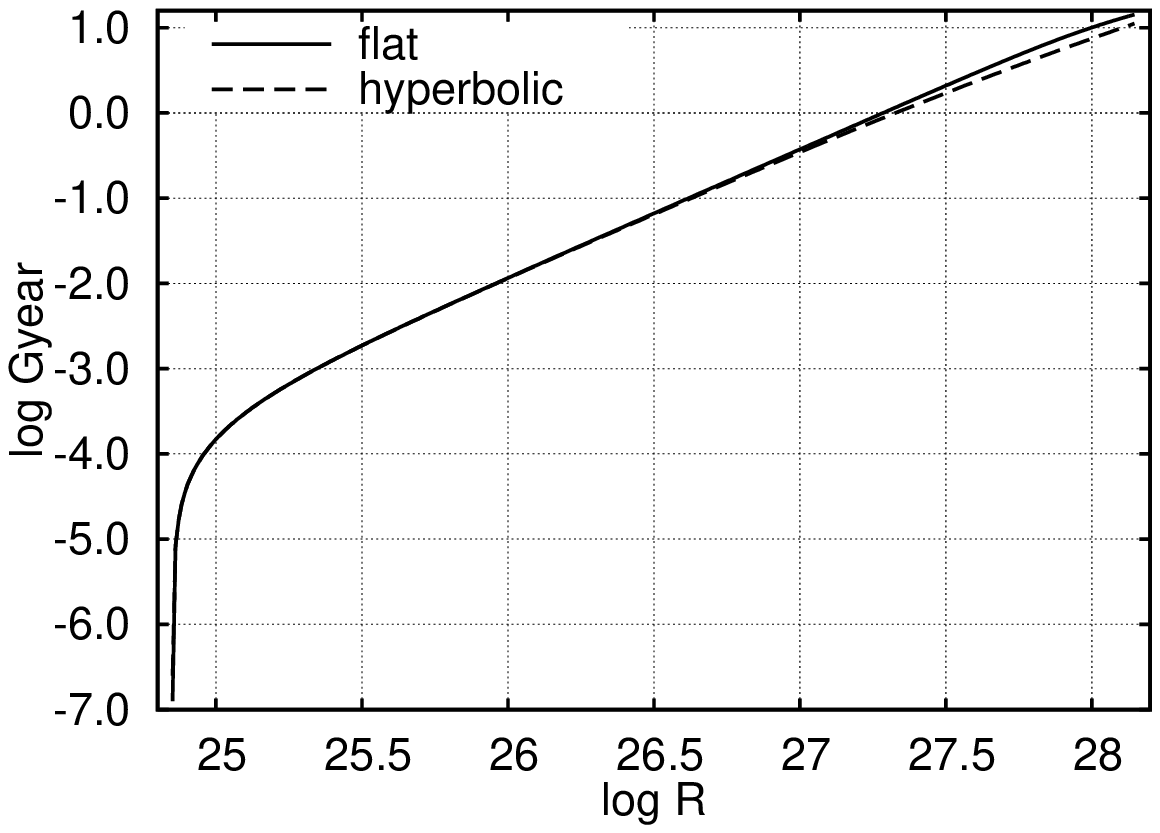,width=100mm,angle=0,clip=}}
\vspace{1mm}
\captionb{2a}
{The age of the universe versus
 $\log R$ for the flat model universe with cosmological constant (solid curve) and for a hyperbolic universe (dashed curve).}
}
\end{figure}
\begin{figure}[!tH]
\vbox{\vspace{-3mm}
\centerline{\psfig{figure=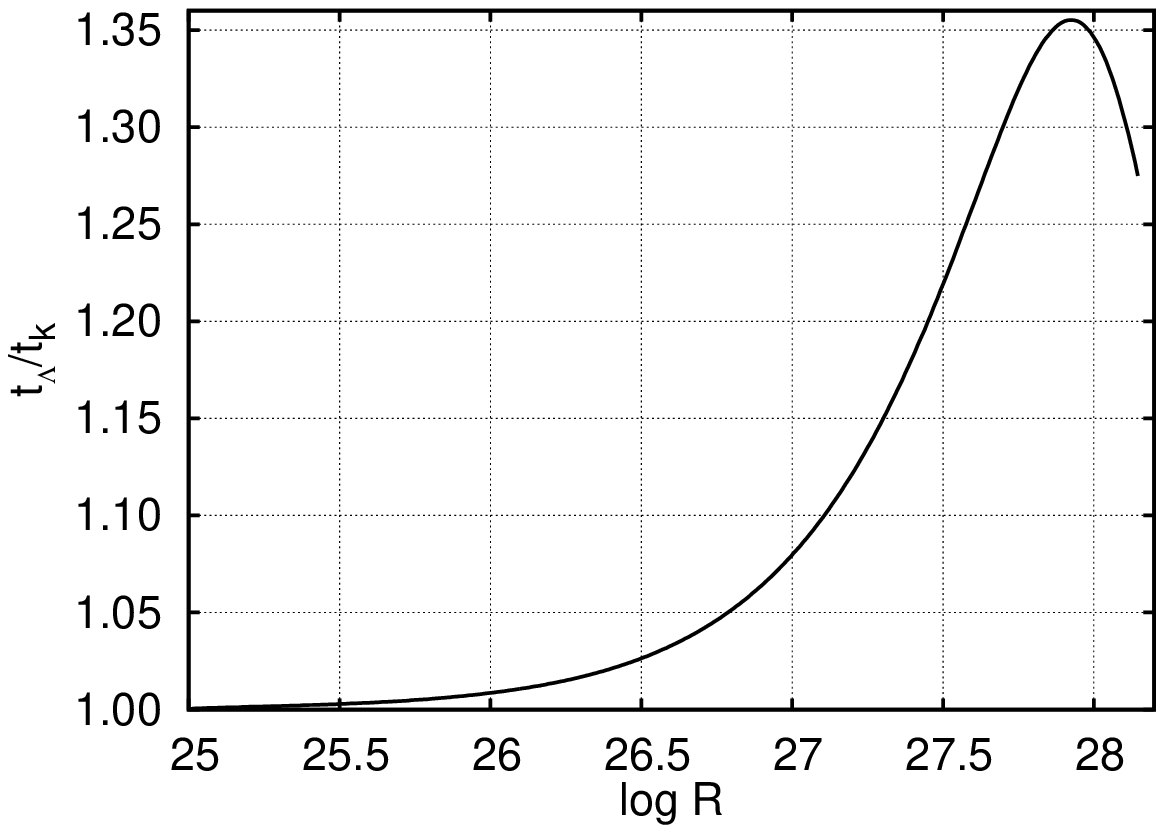,width=100mm,angle=0,clip=}}
\vspace{1mm}
\captionb{2b}
{The ratio of ages of the above described flat and hyperbolic model universes
versus $\log R$.
The difference appears starting at $z<100$.}
}
\end{figure}

As an example, we present here the results of comparison for the flat model universe
with the density of matter $\rho_m=3\cdot 10^{-30}$~g/cm$^3$ and having the cosmological term,
and for a hyperbolic universe having the matter density $\rho_m=2\cdot 10^{-30}$~g/cm$^3$.
For both cases the mass density of the radiation was taken to be $\rho_r=4\cdot 10^{-34}$ g/cm$^3$.
The corresponding  luminosity ratio  in stellar magnitudes, i.e., the expression
 $\Delta m_z=5\log(\omega_\Lambda/\sinh \omega_{-1})$ at $z=1$ is only 0.051.
The respective  curves of the luminosity distance factors $\omega_z$ and $\sinh \omega_z$
are illustrated in Figure\,1a,
and their ratio is given in Figure 1b.

The comparison of the run of the model universe ages is given in  Figure\,2a and their ratio is given in Figure\,2b.
From Fig.\,2a it appears that the evolutionary run of both models is very close.
The  ages of the universes are respectively $t_\Lambda=12.92$~Gyr and
$t_{-1}=11.34$  Gyr. Both the hyperbolic and flat KED universes
presently do not move to the stage of exponential expansion, as in the case of
the presence of the cosmological constant, but are in a transitional stage to the
expansion velocity $\dot R\Rightarrow c$.

It is seen that the difference in the evolutionary run of the universes begins
at $z<10$.
Thus, no essential differences will take place in the formation of
inhomogeneities of the CMB, including the formation of its observed structure
at large $z$ values, when both models almost coincide. However, the duration
of the dark ages and the re-ionization era are somewhat shorter in the case of
a hyperbolic universe. Taking into account the Boomerang mission results of
cosmological triangulation, which state that our universe is at least quite flat, we
conclude that the most appropriate candidate model is the concept of the flat KED
universe.

Let us now make a comment about the comparison of the KED model universe
with the mostly used flat universe with $\Lambda$ generating the impelling force. Whereas
the value of the Hubble constant in the present epoch is well known and the two
first additives with $\alpha$ and $\beta$ are the same in both model universes, we get the
constraint that the third additives must be presently treated as the equal ones,
i.e.
 \begin{equation}
{\Lambda \over 3}={ c^2\over R^2_0}.
\end{equation}
Thus, it can be treated that instead of the cosmological constant we have a
quintessence with the pressure index $-1/3$ which physically corresponds to the
energy integral in the flat expanding universe.

\sectionb{7}{THE EXPANDING BUBBLE UNIVERSE}

In order to overcome the principal difficulty in the Big Bang cosmology, we propose by
{\it Gedankenexperiment}, i.e. an imaginary experiment, the Bubble universe with $k=+1$ and $\kappa=+2$.
In this case all the given equations are valid, the
world is elliptical and of finite volume. In such a model the luminosity distance
has the form
\begin{equation}
D_L=( 1+z)D_g=( 1+z)R_0\sin\omega_z.
\end{equation}
In the Bubble universe the Big Bang really took place locally in a 5-dimensional
Kaluza-Klein type space, namely in a flat special relativity type world (the four
Euclidean spatial coordinates and time), into which our universe can be embedded
(Sapar 1964). In this case Equation (\ref{rz}) is to be replaced by
\begin{equation}
r_z={\sin(\omega_z)\over \omega_z },~~~~~~~~\omega_z=\ln\bigg({y_0\over y_z}\bigg).
\end{equation}
Thus, the luminosity curve of SNe in an elliptical Bubble universe is slightly shifted in the opposite direction than that in a hyperbolic universe, giving a higher ap-
parent brightness.

Taking into account that at the early evolutionary epochs of the universe all
known particles were relativistic, we obtain the equation of evolution for the very
early universe:
\begin{equation}
R^2=2\sqrt{\beta}ct.
\end{equation}
Applying this equation to the pre-Planckian universe, starting from $t=0$,
we obtain that at the Planck time, ${t_P}=(G\hbar/c^5)^{1/2}=5.39\cdot
10^{-44}$~s, the radial coordinate $R$ was about $3.42\cdot 10^{-4}$ cm.
At the Planck time this number exceeds the
Planck length, $l_P=ct_P$, about $10^{30}$ times.
Traditionally, the needed additional
expansion has been ascribed to the inflation of unknown astroparticles -- inflatons.
The inflationary expansion happens somewhat later than the hypothetical
formation of the universe corresponding to the Planck units.

\sectionb{8}{A COMMENT ON THE DARK MATTER}

Neutrino oscillations are a testimony that neutrinos have a small rest mass.
Undoubtedly, neutrinos play in cosmology some role. However, this role is
hitherto not sufficiently understood. Let us study shortly the present state of neutrino
background in cosmology, accepting that the mean rest energy of electron, muon
and tauon type neutrinos is about 1 eV. This means that the mean mass of neutrinos is about $5\cdot10^5$ times smaller than $m_e$.

It is not excluded that the amount of the dark matter is somewhat overestimated,
the mass of tauon neutrinos is not firmly determined and, at last, the right-hand
neutrinos can exist. Therefore, it is not fully excluded that neutrinos
can be one of the main contributors to the dark mass.
Now let us briefly describe one important feature of neutrinos related to the observed
rotation curves of spiral galaxies.

 The decoupled neutrinos on their
non-relativistic stage of energies are cooling according to the law
$ p^2/2m_\nu\propto R^{-2}$. Numerical estimates showed that their
temperature in the present epoch is about $10^{-3}$ K, i.e., their mean
velocity is several hundreds km/sec. Such neutrinos in galaxies move predominantly radially,
and thus they form wide equipotential wells, which can generate the observed flat
rotation velocity curves and accelerate the processes of gravitational clustering
during their formation.

\sectionb{9}{CONCLUSIONS}

The expanding universe can presently be of the KED nature, having dominance
of the kinetic energy with respect to the potential energy. In the mainstream
cosmology, instead of KED, different other possibilities have been studied. We
have modified the traditional approach to the equations of Friedman, introducing
the kinetic energy dominance into the flat universe. In this case the final expansion
rate of the radial coordinate of the universe tends not to infinity, but to the velocity
of light. Thus, the enigmatic dark energy is not more essential in the universe with
a small KED originating from the Big Bang era. The situation can be compared
with a small asymmetry of the matter and antimatter in the universe.

It is not completely excluded that non-relativistic massive neutrinos contribute
to the dark matter background. However, long-term intensive searches of
astroparticles have hitherto remained almost without success. The needed massive
neutrinos with the rest energy 1--3 eV could cool down to very low temperatures, and
they can form the wide equipotential wells in the galaxies, generating the observed
flat velocity rotation curves in the outer regions of galaxies.

 In the current paradigm the hypothetical anti-gravitational particles, the inflatons, are
 tunneling  the energy  from the false vacuum to the real Big Bang. In this way
 the very hot matter  has been created at the conditions of quasi-constant
 extremely hot temperature, whilst the linear dimensions of the universe have increased by
 about 30 dex. If we extrapolate our equation even to the
 zero time considered as the real Big Bang moment,
 then at the Planckian moment the linear dimensions of the
 universe have exceeded the Planckian unit of length by about 30 dex, and
 any  inflationary  epoch in the evolution of the universe is not compulsory.

However, from the theoretical point of view the most elegant is the possibility
of the expanding Bubble universe, in which the creation of the world has been
really a local Big Bang in the 5-dimensional pseudo-Euclidean space, removing
the usual tacitly accepted concept of the infinite space.

Our revised modeling of the universe and its evolutionary scenario are rather
conservative, being based on our aged studies, which have remained almost
unnoticed in the West, as it happens frequently with the papers published in Russian.
Returning to the problem was mainly stimulated by the recent progress in
observational cosmology.

In cosmology the hypothetical and sometimes even paradoxical standpoints
have been always an essential part of any concept. Each new achievement here
removes some difficulties, but generates new puzzling problems. Thus, all versions
of the creation of the universe are somewhat and somehow speculative since they
use extraordinary extrapolations far into the past, being based, however, on the
laws of physics valid here and now. The hope that one of such extrapolations in
its present status is preferable, seems to be a rather overestimated optimism or
ignorance of other possibilities.

To summarize, I have tried to demonstrate that the observational data can be
interpreted without introducing the dark energy and, possibly, the dark matter
of unknown astroparticles. I was encouraged to return to modeling the universe,
which was always among my favorite topics, by an inspiring sentence by Saul
Perlmutter in his Nobel Lecture: ``Everybody talks about the dark energy, but
nobody does anything about it''. For myself, I reformulated it as a business slogan
in science: ``Publish, before you perish''.
\\

 ACKNOWLEDGMENTS. This paper has been supported by the research project
 SF0060030s08 of the Estonian Ministry of Education and Research.

%  \thanks{We are thankful to R. Poolam\"{a}e
%  for his help in preparation of the manuscript.} % to a referee for constructive criticism and

\References

\refb Aldering G., Kim A. G., Kowalski M. et al. 2007, Astroparticle Phys., 27, 213

\refb Astier P., Guy J., Regnault N. et al. 2006, A\&A, 447, 31

\refb Dabrowsky M., Stelmach J. 1986, AJ, 92, 1272

\refb Goldhaber G., Perlmutter S. 1998, Phys. Rep., 307, 325

\refb Kaufman S. E., Schucking E. L. 1971, AJ, 76, 583

\refb Kaufman S.E.  1971, AJ, 76, 751

\refb Leibundgut B. 2009, A\&A, 500, 615

\refb MacTavish C. J., Ade P. A. R., Bock J. J. et al. 2006, ApJ, 647, 799

\refb Sapar A. 1964, Tartu Astr. Obs. Publ., 34, 223

\refb Sapar A. 1965, Tartu Astr. Obs. Teated, 13, 1

\refb Sapar A. 1966, Tartu Astr. Obs. Publ., 35, 368

\refb Sapar A. 1970, AZh, 47, 503

\refb Sapar A. 1976, Tartu Astr. Obs. Publ., 44, 21

\refb Sapar A. 2011, Baltic Astronomy, 20, 267

\end{document}